\documentclass[aps,prl,floats,floatfix,twocolumn,longbibliography,superscriptaddress]{revtex4-1}

\usepackage{color} 
\usepackage{latexsym}
\usepackage{amsmath} 
\usepackage{amssymb} 
\usepackage{bm}
\usepackage{wasysym}

\usepackage{graphicx}
\usepackage{epstopdf}
\graphicspath{{./}{./Figs/}}

\usepackage[colorlinks,linkcolor=blue,urlcolor=blue,bookmarks=true,pdftitle={}]{hyperref}

\newcommand{\mylabel}[1]{\label{#1}} 
\newcommand{\beq}{\begin{eqnarray}}
\newcommand{\eeq}{\end{eqnarray}} 
\newcommand{\be}[1]{\begin{eqnarray}\ifthenelse{#1=-1}
{\nonumber}{\ifthenelse{#1=0}{}{\mylabel{e#1}}}}
\newcommand{\ee}{\end{eqnarray}} 

\newcommand{\hide}[1]{}

\newcommand{\sect}[1]{{\bf #1.-- }}

\newcommand{\Cn}[1]{\begin{center} #1 \end{center}}

\begin{document}

\title{Prethermalization with negative specific heat}

\author{Sayak Ray} 
\affiliation{
Department of Chemistry, Ben-Gurion University of the Negev, Beer-Sheva 84105, Israel}

\author{James R Anglin}
\affiliation{
State Research Center OPTIMAS and Fachbereich Physik, Technische Universit\"at Kaiserslautern, D-67663 Kaiserslautern, Germany}

\author{Amichay Vardi} 
\affiliation{
Department of Chemistry, Ben-Gurion University of the Negev, Beer-Sheva 84105, Israel}

\begin{abstract}
We study non-canonical relaxation in a composite cold atoms system, consisting of subsystems that possess negative microcanonical specific heat. The system exhibits pre-thermalization far away from integrability due to the appearance of a single adiabatic invariant.  The Thirring instability drives the constituent subsystems towards the edges of their allowed energy spectrum, thus greatly enhancing the contrast between the prethermal state and the long time thermal outcome. 
\end{abstract}

\maketitle


The foundations of statistical mechanics are being challenged by demonstrations that not all systems equilibrate canonically. Dynamically integrable systems may evolve towards non-canonical steady states \cite{int1,int2,int3,int4} because without chaos systems do not ergodically explore energy shells. While truly integrable systems are rare, ``nearly'' integrable systems may remain in non-canonical steady states over significant time scales, a phenomenon known as  \textit{prethermalization} \cite{int3,int4,pretherm1,pretherm2,Marcuzzi13,pretherm3,pretherm4,pretherm5,pretherm6,Lerose19}. 

It has recently been pointed out that prethermalization does not necessarily require near-integrability \cite{Lenarcic18,Lange18,Mallayya19}. When dynamics conserves additional motional constants besides energy and particle number, the system relaxes towards an equilibrium described by a generalized Gibbs ensemble (GGE) instead of the standard grand canonical ensemble (GCE). This is true even if the total number of constants of motion is much smaller than the number of degrees of freedom. If a small perturbation then breaks these additional conservation laws, the system will still initially relax into the same GGE as a 'prethermal state', but then relax further into the standard GCE as it slowly migrates out of the reduced phase space shell that was defined by the no-longer-conserved constants.

In this work we present an example of such prethermalization in a model system which can be realized experimentally with quantum gases and in which all the important thermodynamical properties can be computed analytically. Non-canonical equilibration into a GGE occurs here not due to an extra exact symmetry but because of an additional adiabatic invariant; prethermalization occurs as the adiabatically approximate invariance breaks down. The dramatically non-canonical nature of equilibration in this case is related to the fact that the component subsystems of this aggregate system all have {\em negative specific heat} (NSH). 

\sect{Negative specific heat}
The Second Law of Thermodynamics usually requires heat in an aggregate system to disperse among its constituent subsystems until equilibrium at uniform temperature is reached, but this changes when the heat capacity $C = d\varepsilon/dT$ of each subsystem---the \emph{specific heat}, in terms of the subsystem energy $\varepsilon$---is negative. Negative specific heat (NSH) implies Thirring instability\cite{Thirring70}: a subsystem which absorbs heat from its surroundings because it is colder will thereby become colder still, and conversely a hotter subsystem will be heated further by losing energy. No violation of the Second Law is involved, since heat always flows from hotter systems to colder and entropy $S$ only increases, yet heat spontaneously \emph{concentrates}.

The most important examples of negative specific heat are in self-gravitating astrophysical systems \cite{Eddington26, Schwarzschild58, Thirring70, Hertel71, Bell99}, such as cold cosmic gas clouds in which spontaneous hot spots become stars, or protoplanetary discs in which planets heat by accretion. Negative specific heat is also observed in various other long-range interacting systems \cite{Campa09}, clusters of atoms and molecules \cite{Schmidt01, Bixon89, Labastie90}, and in the fragmentation of nuclei \cite{Agostino00}. 

NSH is thus real but its meaning in statistical mechanics is subtle. Thermodynamics with NSH is inherently non-extensive \cite{Thirring70}, so that an aggregate of NSH subsystems typically has positive $C$. Moreover only the microcanonical $C$ can be negative. Each subsystem in an aggregate is automatically coupled to a heat bath composed of all the other subsystems, however, and so the ensemble of NSH subsystems will have a probability distribution of fluctuating $\varepsilon$, rather than a microcanonically definite $\varepsilon$, even if the aggregate's total energy $E$ is fixed. 

Microcanonical NSH nevertheless remains a real property of the subsystems and can have dramatic effects. The Thirring instability implies bistability, where each subsystem tends to have either much lower or much higher $\varepsilon$ than the aggregate average: heat concentrates rather than dispersing, and steady states are not uniform. The distribution of $\varepsilon$ over the aggregate is well-defined but \emph{bimodal}. Thirring's thermodynamic explanation of the instability, based on microcanonical subsystem temperature $T=1/S'(\varepsilon)$ \cite{Thirring70}, translates straightforwardly into canonical language: microcanonical $C=-S''(\varepsilon)/S'(\varepsilon)^2<0$ implies an anomalously high density of states at high energies, providing a second probability peak there even when a low canonical bath temperature $T_{B}$ otherwise favors low energy. 

\begin{figure}[t]
\centering
\includegraphics[clip=true,width =1\columnwidth]{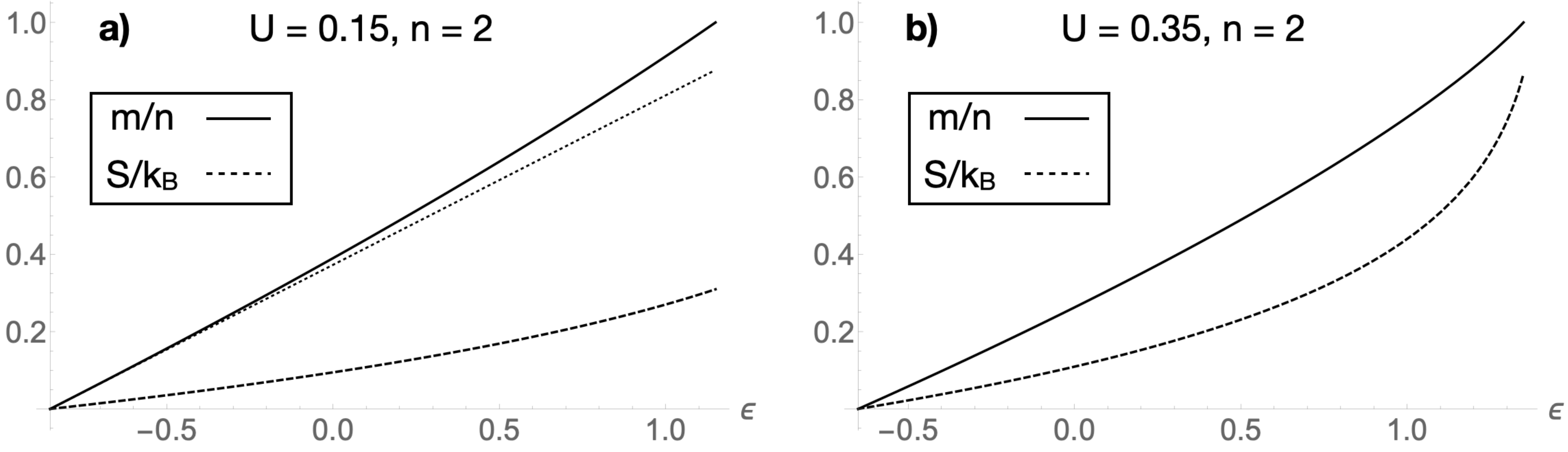}
\caption{The dimensionless action variable $m(\varepsilon)$ (solid) and microcanonical entropy $S(\varepsilon)/k_B = \ln[m'(\varepsilon)/m'(\varepsilon_-)]$ (dashed) of the Bose-Hubbard dimer as a function of dimer energy $\varepsilon$ for $n=2$ and $U=0.15$ (a) or $U=0.35$ (b). The dotted straight line in (a) is to help show the upward curvature of $m$. The curves are representative of all $0<Un<1$; upward curvatures increase with $Un$. The slope of $S$ is everywhere positive, meaning positive microcanonical temperature $T=1/S'(\varepsilon)$. As implied by the monotonically increasing slope of the $S(\varepsilon)$ curve, the microcanonical specific heat $C=-S''/(S')^2$ of a dimer is negative for all $\varepsilon$ for $0<Un<1$.}
\label{f1}
\end{figure} 

\sect{Bose-Hubbard model system}
To test whether such unusual bimodal distributions can really emerge in the time evolution of an isolated non-integrable dynamical system, we consider a large $L\times L$ two-dimensional array of identical two-mode Bose-Hubbard (BH) subsystems with repulsive on-site interactions, weakly linked to each other by nearest-neighbor tunneling. Realizable as the tight-binding limit of dilute ultracold bosons trapped in a lattice potential, this quantum system is represented accurately for large particle numbers by the semi-classical mean-field Hamiltonian
\begin{eqnarray}
\mathcal{H} &=& \sum_{r} H_{r} -\frac{J}{2}\sum_{\langle r, r^{\prime}\rangle} \sum_{\sigma=1}^{2}(\alpha^{*}_{\sigma,r} \alpha_{\sigma,r^{\prime}} + \text{c.c.}), \nonumber \\ 
H_{r} &=& -\frac{\Omega}{2}(\alpha^{*}_{1,r} \alpha_{2,r} + \text{c.c.}) + \frac{U}{2}\sum_{\sigma=1}^{2}n_{\sigma,r}^2 \label{BHM_2D}
\end{eqnarray} 
where $H_{r}$ is the two-mode BH (``dimer'') Hamiltonian \cite{Chuchem10} at each site $r\equiv (i,j)$ of the 2D lattice, with onsite interaction $U$ and coupling $\Omega$ between the two modes $(\sigma=1,2)$ at each site. The complex amplitudes $\alpha_{\sigma,r}$ are mean-field representations of second-quantized bosonic destruction operators, so that $n_{\sigma,r}\equiv|\alpha_{\sigma,r}|^{2}$ is a particle number represented in mean-field approximaton as continuous. Inter-dimer coupling is provided by tunneling with rate $J$ between nearest neighboring dimers $\langle r, r^{\prime}\rangle$. The total particle number $N=\sum_{\sigma,r}n_{\sigma,r}$ is conserved, as is the value $E$ of the aggregate Hamiltonian $\mathcal{H}$. In (\ref{BHM_2D}) we have already set $\hbar=1$; throughout this paper we will generally also set $\Omega=1$.

If coupling between dimers is neglected then each single dimer is integrable, with action-angle variables that can be constructed analytically. We will focus in this paper exclusively on the regime $0<U n <1$ in which the individual dimers exhibit Josephson oscillations without self-trapping. We assume $J\ll \Omega$ ($J\ll1$) in order to implement weak coupling between the subsystems, so that $E\doteq \sum_r{\varepsilon}_{r}$ where $\varepsilon_{r}$ is the value of $H_{r}$. Since we consider time scales much longer than the inter-dimer tunneling time, however, our large aggregate system is \emph{not} effectively decoupled and integrable. Its only exact constants of the motion are $N$ and $E$.

\sect{Subsystem negative specific heat}
To derive the thermodynamical properties of the one-dimer subsystems we examine $H_r$ for a single dimer, dropping the $r$ subscripts. Since $n=n_1+n_2$ is conserved the phase space is effectively two-dimensional, spanned by $n_1-n_2$ and the relative phase of $\alpha_{1,2}$; single-dimer properties only depend on $U$ through the product $u=Un$. The energy $\varepsilon$ is bounded from above and below by $\varepsilon_{\pm}=\pm n/2 + un/4$ (for $u<1$). The two-dimensional phase space area  enclosed between a contour of fixed energy $\varepsilon$ and the ground state, which is 2$\pi$ times the action coordinate $m(\varepsilon,u)$, can be expressed exactly in terms of complete elliptic integrals (see our Supplementary Material \cite{sup}), and this yields further analytic expressions for the microcanonical entropy $S=k_B\ln(2\pi\partial_\varepsilon m)$, temperature $T=1/\partial_\varepsilon S$, and specific heat $C=1/\partial_\varepsilon T$. As previously noted in \cite{Strzys14} and illustrated in Fig.~1, $T$ is positive and $C$ is negative for all $\varepsilon_-<\varepsilon<\varepsilon_+$ and all $0<u<1$. Do we see Thirring bimodality in the distribution of dimer energies after the whole array has evolved under $\mathcal{H}$ for long times?

\begin{figure}[t]
\centering
\includegraphics[clip=true,width =1\columnwidth]{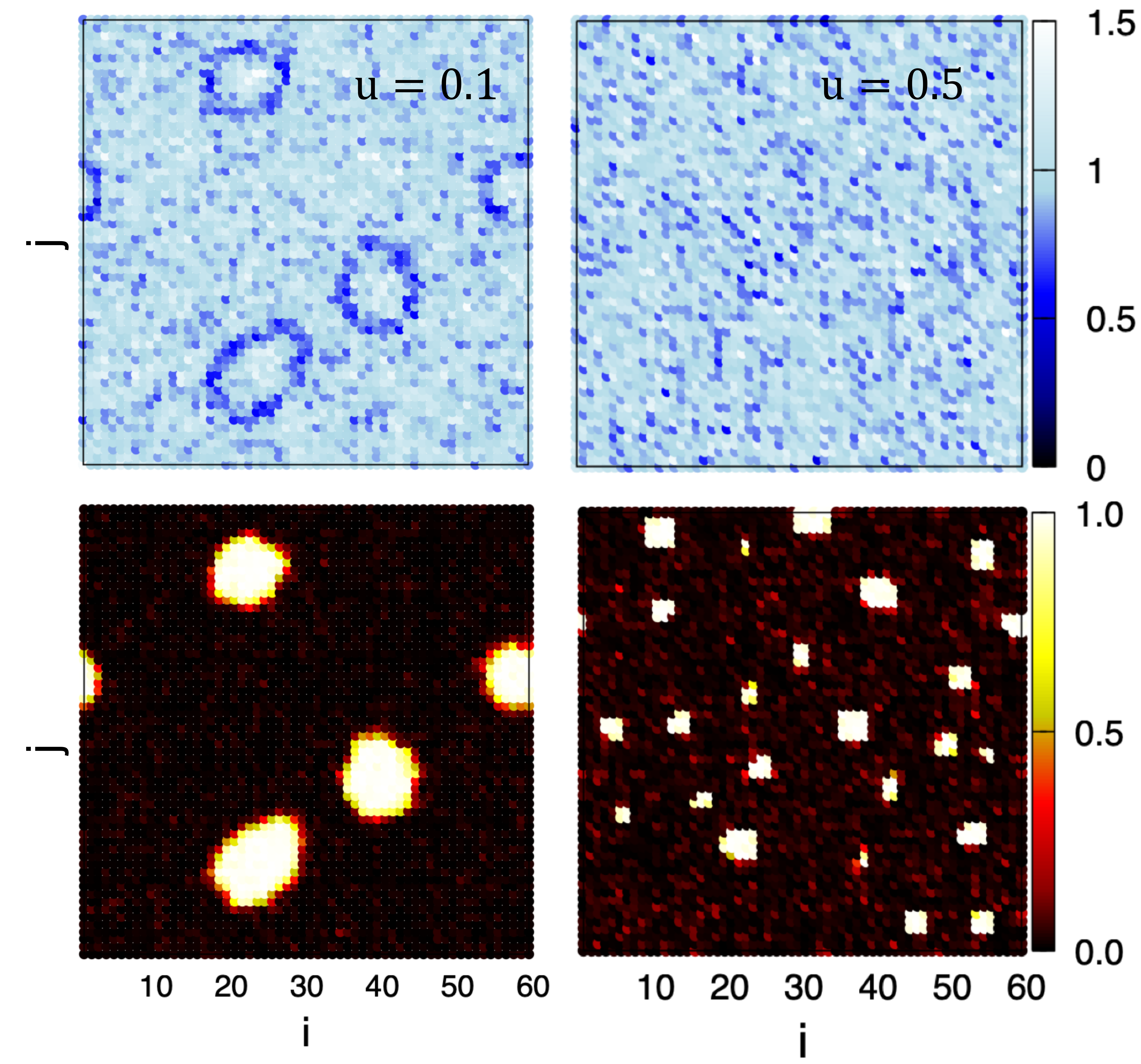}
\caption{(Color online) Long-time numerical propagation: The Bose-Hubbard dimer populations $n_{ij}$  (a,b) and energies $\varepsilon_{ij}$ (c,d) at $Jt=10^3$ are plotted throughout the $60\times60$ dimers array. Initial conditions correspond to a uniform population and energy distribution (see black dot in Fig.~\ref{f3}); panels a,c  and b,d show the steady state outcome at $u=0.1$ and $u=0.5$, respectively. In all simulations $J=0.05\Omega$.}
\label{f2}
\end{figure} 

\sect{Numerical results}
As Fig.~\ref{f2} shows, we do. In Fig.~\ref{f2} we plot all the single dimer energies $\varepsilon_{r}$ and occupation numbers $n_r$ of the whole array, after a long time evolution, from generic initial conditions with no particular symmetry. The structures seen in the Figure form spontaneously. In the very small-$U$ limit, the inter-dimer coupling $J$ is sufficiently competitive to produce significant surface tension in the domain walls that surround energy concentrations, making them behave as mobile droplets \cite{Strzys14}. For stronger interaction, the surface tension becomes negligible, favoring small immobile energy breathers \cite{Dey17}. Equally important from a thermodynamic point of view are the dark voids between the bright spots, where dimers are all low in energy. No effort to simulate NSH has been made in this Figure; the results emerge purely from Hamiltonian evolution, just as they should if entropy increase accurately represents ergodic evolution in a non-integrable system.

\begin{figure}[t]
\centering
\includegraphics[clip=true,width =1\columnwidth]{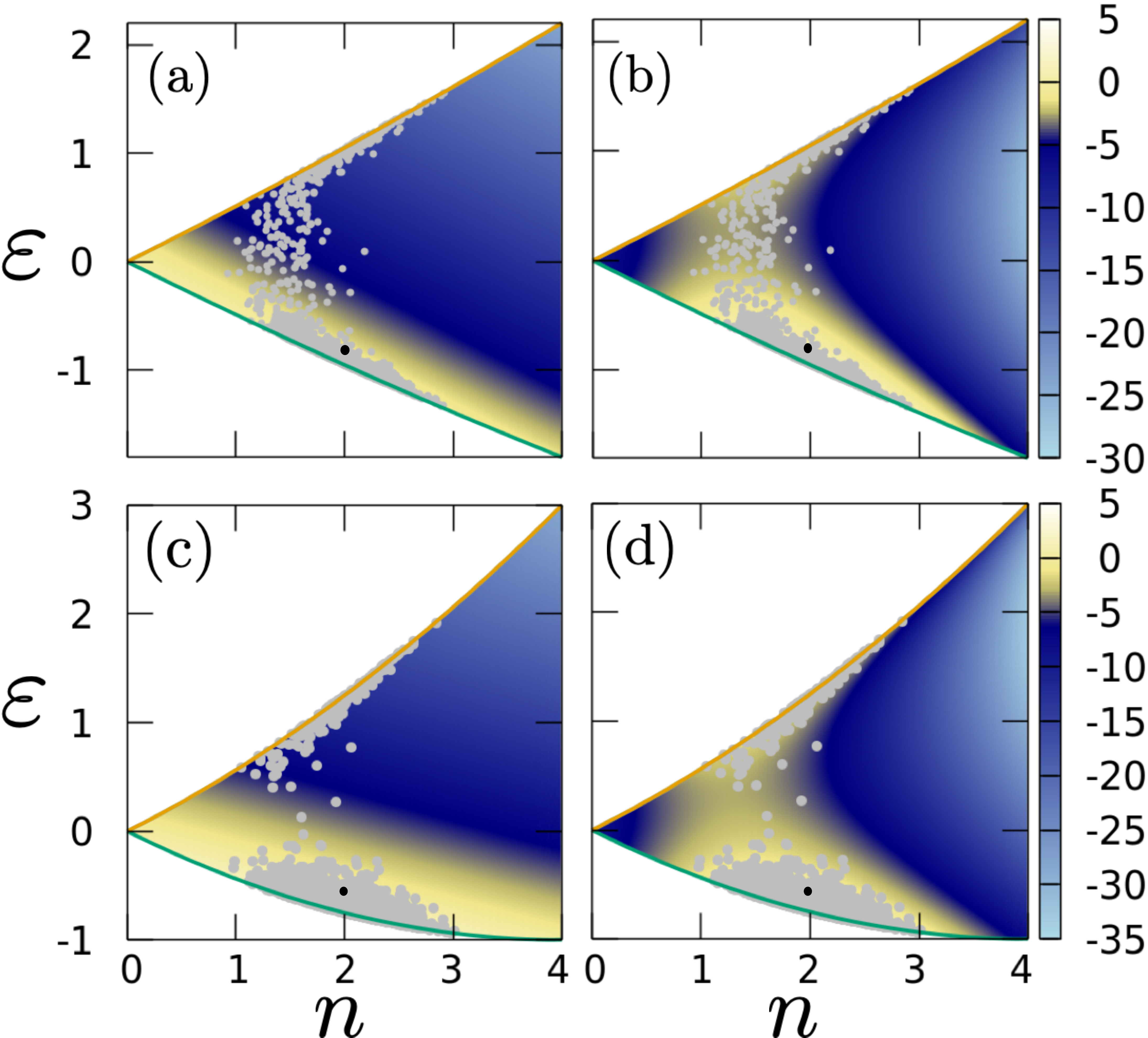}
\caption{(Color online) Comparison of the one-dimer number and energy distribution in the results of Fig.~\ref{f2} (gray dots, each corresponding to a single dimer) with the GCE
distribution of Eq.~\ref{GCE} (a,c for $u=0.1$ and $u=0.5$ respecively) and with the generalized Gibbs distribution of Eq.~\ref{GGE} (b,d for the same parameters). The black dot corresponds to the initial uniform distribution. While the GCE fails to describe the observed distribution, the GGE captures the numerical outcome, thus demonstrating restricted ergodicity within the $N,E,M$ shell.}
\label{f3}
\end{figure}

\sect{Non-canonical thermalization} 
Careful analysis shows that there is more than NSH going on in this system, however. In the presence of microcanonical NSH a GCE can be bimodal in energy, but in fact the ensemble of all the dimers in our array cannot be described by a GCE. In Fig.~\ref{f3}a,c we compare the $n,\varepsilon$ distribution of the dimer subsystems in Fig.~\ref{f2} to the GCE distribution
 \begin{equation}
 P_{\rm GCE}(n,\varepsilon)=\frac{\partial m(\varepsilon,u)}{\partial \varepsilon}\frac{e^{\frac{\varepsilon-\mu n}{k_B T_B}}}{{\cal Z}(T_B,\mu)}
 \label{GCE}
 \end{equation}
where ${\cal Z}(T_B,\mu)$ is the grand partition function, with $T_B$ and the chemical potential $\mu$ set by the constraints
\begin{equation}
\sum_{n,\varepsilon} \varepsilon P(n,\varepsilon)=\frac{E}{L^2}~,~\sum_{n,\varepsilon} n P(n,\varepsilon)=\frac{N}{L^2}~.
\label{GCC}
\end{equation}
It is clear that no GCE can describe the steady state that our BH array has reached. The reason is that the BH dimer energy $\varepsilon$ is bounded from above by an upper limit $\varepsilon_{+}(U,n)$, and since it has NSH, it has a \emph{minimum} microcanonical temperature $T_{-}=T(\varepsilon_{+})$. The array energy in the case shown in Fig.~\ref{f3} implies a canonical bath temperature $T_B$ which is \emph{less than} $T_{-}$, so in this case no dimers can be colder than their collective bath.  

So we are seeing Thirring-like bimodality when NSH alone cannot provide it. And in fact we see a much more dramatic bimodality, with many dimers clustered tightly around both $\varepsilon_\pm$ and very few in between, than the only slightly bimodal, rather flat distribution that the modest convexity of our microcanonical entropy would predict even in the right range of $T_B$. Why does this happen?


\sect{Adiabatic invariance of the single-dimer action variable} The coupling $J$ between the dimers is not simply negligible. Over the long evolutions that we follow it has ample time to affect the system decisively. Because $J\ll\Omega$, however, there can be a large separation of time scales between the intra-dimer and inter-dimer dynamics. Below we will discuss exactly when this time scale hierarchy exists, but when it does the sum of all the action variables of the individual dimers $M=\sum_r m_r$ is an \emph{adiabatic invariant} \cite{Strzys10,Strzys12,Strzys14}. Adiabatic invariance is not exact conservation; an adiabatic invariant has, as it were, weather but no climate change. The secular trend in an adiabatic invariant is zero exactly. 

For thermodynamics and statistical mechanics of our weakly coupled Bose-Hubbard array, therefore, the adiabatic invariance of $M$ is as good as exact conservation. The whole array as a Hamiltonian system does not ergodically explore the entire phase space shell specified by $E,N$, but is restricted instead to the $E,N,M$ subspace when $M$ is adiabatic invariant. When we consider each dimer as a system coupled to a bath composed of all the other dimers, therefore, we must consider that bath and system can exchange not just two quantities that are in total conserved, but three: $E$, $N$, and $M$.

\sect{The generalized Gibbs ensemble} For the ensemble of individual dimers we therefore construct a GGE in the form:
 \begin{equation}
 P_{\rm GGE}(n,\varepsilon)=\frac{\partial m(\varepsilon,u)}{\partial \varepsilon}\frac{e^{\frac{\varepsilon-\mu n-{\tilde \mu}m(n,\varepsilon)}{k_B T_B}}}{{\cal Z}_{\rm GGE}(T_B,\mu,\tilde\mu)}
 \label{GGE}
 \end{equation}
where ${\cal Z}_{\rm GGE}(T_B,\mu,\tilde\mu)$ is the GGE partition function and the added constraint,
\begin{equation}
\sum_{n,\varepsilon} m(n,\varepsilon) P(n,\varepsilon)=\frac{M}{L^2}~,
\label{GGC}
\end{equation}
is used along with Eqs.~(\ref{GCC}) to set the values of the Lagrange multipliers $T_B,\mu$ and $\tilde\mu$.
This generalized Gibbs distribution is shown in Fig.~\ref{f3}b,d. Unlike the GCE distribution it is sharply bimodal, agreeing well with the long-time Hamiltonian evolution of the dimer array and thus confirming that Gibbs ensembles represent the ergodicity of non-integrable Hamiltonian evolution. 

The bimodality of the GGE is easy to recognize from Fig.~1: $m(\varepsilon)$ is a nearly straight line that curves slightly upwards. It is therefore easy to tune $\tilde{\mu}$ to make the GGE's Boltzmann exponent $-(\varepsilon-\tilde\mu m)/(k_{B}T_{B})$ U-shaped. This effect is similar to that of the convexity of the microcanonical entropy $S$, which is just the logarithm of the density of states prefactor $\partial m/\partial{\varepsilon}$ in both probabilities, but because the Boltzmann factor also contains $T_{B}$ the bimodality of the GGE due to $\tilde{\mu}m$ can be arbitrarily strong for low enough temperatures. Since the microcanonical $S$ is simply the log of the derivative of $m$, the upward curvature of $m$ that drives this stronger bimodality is directly related to the upward curvature of $S$ that defines NSH. In effect the adiabatic invariance of $M$ amplifies the system's basic NSH character, and Thirring instability due to the density of available states makes equilibration in this system dramatically non-canonical.

\begin{figure}[t]
\centering
\includegraphics[clip=true,width =0.98\columnwidth]{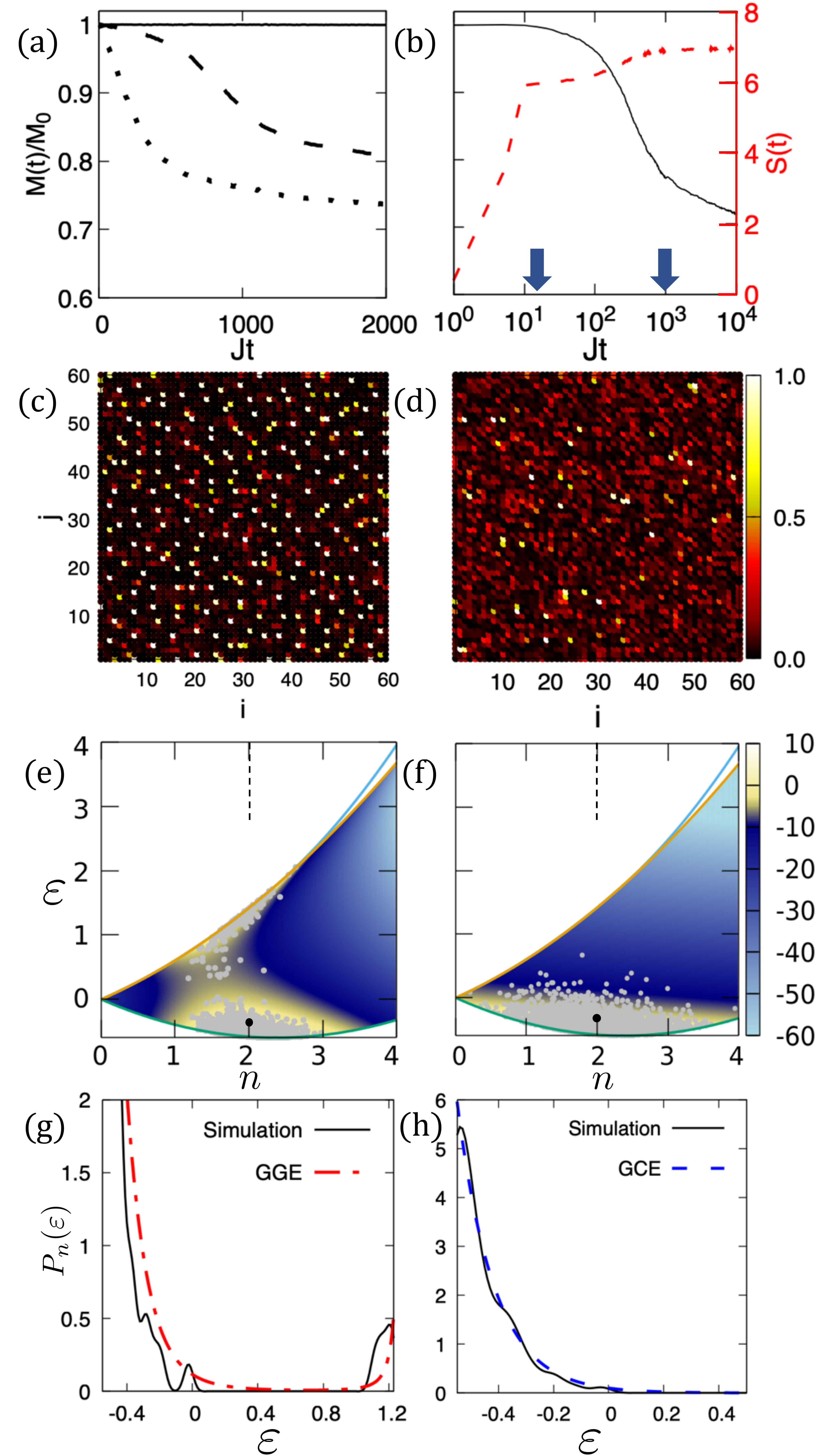}
\caption{(Color online) Prethermalization: (a) Time evolution of the total joson number $M$ for $u=0.5$ (solid), $0.7$ (dashed) and $0.9$ (dotted) and $J=0.05\Omega$; (b) Joson number and total entropy dynamics for $\bar{u}=0.84$, just above the critical value $u_c=0.8$. Arrows mark the times at which panels c,e,g and d,f,h are taken; (c) The $\varepsilon_{r}$ distribution at the end of the M-conserving prethermalization stage; (d) same after complete thermalization; (e) Prethermal state matches the GGE with the appropriate $N,M,E$; (f) Thermal state matches GCE with the same $N,E$. Panels g,h compares crosscuts through the GGE and GCE distributions at the most probable value of $n$ (marked by the dashed line in e,f) showing good agreement with the numerical result at the corresponding stage.}
\label{f4}
\end{figure} 

\sect{Prethermalization without near-integrability} The adiabatic invariance of $M$ does not follow merely from $J\ll\Omega$ but from the inter-dimer tunneling time $1/J$ being much longer than the single-dimer dynamical time scale, which depends on $u$ and $\varepsilon$ \cite{Strzys10,Strzys12}. For $u\to1$ and $\varepsilon\to\varepsilon_+$ this time scale goes to zero, as large-amplitude nonlinear Josephson oscillations become slow in the approach to self-trapping. For any finite $J$ there is therefore a threshold $u$ above which the invariance of $M$ breaks down. This threshold can be computed by Bogoliubov stability analysis of the array about the maximally excited state with all $n_{r}=N/L^2=n$ and all $\varepsilon_{r}=\varepsilon_{+}(n)$ \cite{sup}. This identifies a modulational instability for $u> u_c=1-4J$. Indeed, when we calculate $M$ throughout the time evolution in Fig.~\ref{f4}a, we see it is conserved for $UN/L^{2}<u_c$, but at higher values of average $u$, $M$ decays over time.

Repeating our time evolution at a supercritical value of $UN/L^{2}$, in Fig.~4 we observe two clear stages. In the faster prethermalization stage $M$ is conserved and the system explores the restricted shell defined by the initial $N,E,M$, so that the dimer ensemble fits the bimodal GGE. Over longer times, however, $M$ is no longer free of secular trends; the system gradually migrates out of the initial $M$ shell to explore the full $N,E$ phase space. At late times the dimer ensemble fits the usual GCE. This is thus a classic case of prethermalization even though near-integrability in this system concerns only the single quantity $M$, while the $2L^2-3$ other degrees of freedom in the array are all far from integrable.

\sect{Conclusions} Considering the thermalization of a homogeneous aggregate system consisting of subsystems with negative microcanonical specific heat, we have shown that the system approaches non-canonical distributions even though it is extremely non-integrable. A single extra adiabatic invariant, not even exactly conserved, requires a generalized Gibbs ensemble. The breakdown of this adiabatic invariant above a parameter threshold introduces slow phase space migration manifesting as prethermalization. These effects are dramatically enhanced in this system, showing up in sharply bimodal energy distributions, due to the underlying presence of negative specific heat. All of this unusual behavior is explained in terms of analytically computable quantities, supporting the basic assumption of statistical mechanics that maximum entropy ensembles represent ergodicity, even in exotic cases where the subspaces which are ergodic must be considered carefully. Explicit analytical formulas for all microcanonical thermodynamical properties of the two-mode Bose-Hubbard model are given in the Supplementary Material \cite{sup}.

\sect {Acknowledgements}
This work was supported by the Israel Science Foundation (Grant No. 283/18) and by the Deutsche Forschungsgemeinschaft (DFG) in Project number 277625399, TRR 185.


\clearpage
\onecolumngrid
\pagestyle{empty}
\renewcommand{\thefigure}{S\arabic{figure}}
\setcounter{figure}{0}
\setcounter{equation}{0}

\Cn{{\LARGE SUPPLEMENTARY MATERIAL}}

\section{Action variable for the classical two-mode Bose-Hubbard model with $0 \leq u < 1$}

We consider the single-dimer Hamiltonian $H_r$ in Eq.\ (3), with particle number $n_r=|\alpha_{1r}|^2+|\alpha_{2r}|^2$. We henceforth drop the $r$ subscripts, so that we begin from
\begin{equation}\label{Hsup1}
H = -\frac{1}{2}(\alpha^*_1\alpha_2 + \alpha^*_2\alpha_1) +\frac{U}{4}n^2+\frac{U}{4}(n_1-n_2)^2\;.
\end{equation}

As common in studies of the two-mode Bose-Hubbard model we next transform to the Schwinger representation of angular momentum,
\begin{align}
\frac{\alpha_1^*\alpha_2+\alpha_2^*\alpha_1}{2} &= L_1\nonumber\\
\frac{\alpha_1^*\alpha_2-\alpha_2^*\alpha_1}{2i} &=L_2\nonumber\\
\frac{|\alpha_1|^2-|\alpha_2|^2}{2} &=L_3\;.
\end{align}
Although more commonly introduced quantum mechanically, where it is justified by the fact that the canonical commutation relations of the bosonic creation and destruction operators reproduce the angular momentum algebra, this transformation also works classically, with the classical Poisson brackets replacing the quantum commutators. We further note that this representation implies $L_1^2+L_2^2+L_3^2\equiv (n/2)^2$. With this transformation we render (\ref{Hsup1}) into
\begin{equation}
H = -L_1 +U L_3^2+ \frac{Un^2}{4}\;.
\end{equation}

We then introduce a canonical representation of the angular momentum variables for fixed $n$:
\begin{align}
L_1 &= P\nonumber\\
L_2 &= \sqrt{(n/2)^2-P^2}\cos\phi\nonumber\\
L_3 &= \sqrt{(n/2)^2-P^2}\sin\phi\;,
\end{align}
where $\phi,P$ are a canonically conjugate pair with the finite ranges $-n/2\leq P \leq n/2$ and $-\pi < \phi \leq \pi$, $\phi =\pm\pi$ being identified so that we have a finite cylindrical phase space. (It is easy to confirm that the canonical Poisson brackets of $\phi$ and $P$ reproduce those of the angular momentum variables.) This delivers
\begin{equation}
H = -P + U[(n/2)^2-P^2]\sin^2\!\phi + Un^2/4\;.
\end{equation}
A key feature of this our final transformation is that for $0<Un<1$ all the contours of constant $H=\varepsilon$ sweep through the entire $[0,2\pi]$ range of $\phi$. 

For $0<Un<1$ it can also be shown that $Un^2/4-n/2\leq\varepsilon Un^2/4+n/2$.  We therefore represent the energy in terms of the angle $\eta\in [-\pi/2,\pi/2]$ according to
\begin{equation}
\varepsilon = \frac{Un^2}{4} + \frac{n}{2}\sin\eta
\end{equation}
and express the contour of constant energy $H=\varepsilon$ in the $(\phi,P)$ phase space as
\begin{equation}
U\sin^2\phi\, P^2 + P + \frac{n}{2}\sin\eta - \frac{Un^2}{4}\sin^2\!\phi= 0\;.
\end{equation}
Solving the quadratic equation, and discarding one branch because we must have $P\geq -n/2$, gives
\begin{align}
P(\phi,\eta,n) &= \frac{1}{2U\sin^2\phi}\left(-1 + \sqrt{1 - 2Un\sin\eta\,\sin^2\phi + (Un)^2\sin^4\phi}\right)\nonumber\\
&\equiv\frac{n}{2u\sin^2\phi}\left(-1 + \sqrt{1 - 2u\sin\eta\,\sin^2\phi + u^2\sin^4\phi}\right)\equiv n X(\phi,\eta,u)\;,
\end{align}
for $u=Un$. 

The area enclosed for given $\eta$ between this contour $-n/2\leq P(\eta,\phi)\leq n/2$ and the minimum energy contour $P(-\pi/2,\phi)=n/2$ is then by definition equal to $2\pi$ times our action variable $m(\varepsilon,n)$:
\begin{align}
m(\varepsilon,n) &= \frac{n}{2} - \oint\!\frac{d\phi}{2\pi}\,P\Big(\phi,\eta(\varepsilon,n),n\Big)\nonumber\\
& = n\left(\frac{1}{2}-\oint\!\frac{d\phi}{2\pi}\,X\Big(\phi,\eta(\varepsilon,n),u\Big)\right)\;.
\end{align}

Defining the new integration variable $z=\sin^2\phi$ and then integrating by parts allows us to write $m(\varepsilon,n)$ as
\begin{equation}
m = n\left(\frac{1}{2}- \frac{1}{\pi} \int^{1}_{0} dz \sqrt{\frac{1-z}{z}} \frac{uz-\sin \eta}{\sqrt{(1+iue^{i\eta}z)(1-iue^{i\eta}z)}} \right)\;.
\end{equation}
The integral over $z$ is in the general class of elliptic integrals but it does not reduce immediately to any of the three standard Legendre forms of elliptic integral. Mathematica evaluates it exactly in terms of Appell $F_1$ functions, giving a form which is manifestly real but unfamiliar and also slow to compute numerically. If instead of taking $z=\sin^2\phi$ we change variables to $\xi=\cot\phi$ we can obtain an exact evaluation in terms of incomplete elliptic integrals at infinite argument. Applying elliptic integral identities we can then reduce these to complete elliptic integrals of the three basic Legendre types, and confirm by numerical plotting that the result is exactly equal to the exact result in Appell functions. Most of our analytical results in this paper are based on this evaluation:
\begin{align}
\frac{m}{n} =&\frac{1}{2} - \frac{1}{\pi}\mathrm{Re}\left[ 2\cos\eta\frac{K\!\!\left(\frac{1-i e^{-i \eta } u}{1+ie^{i \eta }u}\right)}{\sqrt{1+i e^{i \eta } u}}+\frac{2i}{u}\sqrt{1+i e^{i \eta } u}\, E\!\!\left(\frac{1-i e^{-i \eta} u}{1+ie^{i \eta }u}\right)
 +\frac{i e^{i \eta } \left(1-i e^{-i \eta } u\right) \Pi\!\!\left(i e^{-i \eta } u|\frac{2 i u \cos  \eta }{1+ie^{i \eta } u}\right)}{\sqrt{1+i e^{i \eta } u}}\right]\;,
\end{align}
where $K(m)$, $E(m)$, and $\Pi(n|m)$ are respectively the complete elliptic integrals of the first, second, and third kinds. 

This final expression is admittedly ponderous but it evaluates efficiently numerically and the complete elliptic integrals obey a number of well-known identities, so that derivatives of $m$ with respect to $\varepsilon$ can also all be evaluated exactly in terms of complete elliptic integrals. In particular the microcanonical entropy, temperature, and specific heat can all be obtained straightforwardly, because through further elliptic integral identities the derivative of $m$ with respect to $\varepsilon$ simplifies dramatically to just
\begin{equation}
\frac{\partial m}{\partial\varepsilon} = \frac{2}{\pi}
\frac{K\!\!\left(
	1-\frac{1-i e^{-i \eta} u} {1+ie^{i\eta}u}
	\right)}
	{\sqrt{1+i e^{i\eta}u}}\;,
\end{equation}
which is real in spite of its complex arguments because of an identity satisfied by complete elliptic integrals of the first kind. 

Since $2\pi m$ is the action variable for the Bose-Hubbard dimer, $2\pi m'(\varepsilon)$ not only gives the quantum density of states in the semiclassical limit, and therefore the microcanonical entropy, but is also the period of the classical orbit at energy $\varepsilon$. $K(0)\equiv\pi/2$, so we can confirm that
\begin{equation}
\lim_{\varepsilon\to\varepsilon_\pm} \frac{\partial m}{\partial\varepsilon} = \lim_{\eta\to\pm\pi/2}\frac{\partial m}{\partial\varepsilon} = \frac{1}{\sqrt{1\mp u}}\;,
\end{equation}
which agree with the inverse Josephson frequencies of small oscillations around the ground and highest excited states.

The microcanonical temperature can be expressed in terms of complete elliptic integrals as well:
\begin{equation}
\frac{1}{k_BT} = \frac{u-\sin\eta}{2n\cos^2\eta (1-i e^{-i\eta}u)}\frac{E\!\!\left(
	1-\frac{1-i e^{-i \eta} u} {1+ie^{i\eta}u}\right)}{K\!\!\left(
	1-\frac{1-i e^{-i \eta} u} {1+ie^{i\eta}u}\right)}+\frac{\sin\eta}{2n\cos^2\eta}\;,
	\end{equation}
which is again identically real in spite of its complex arguments. The temperature is also finite at minimum and maximum energy in spite of the $1/\cos^2\eta$ factors, and finite for $n\to0$ as long as $U>0$. The limits are
\begin{equation}
k_B T_\pm = k_B T(\varepsilon_\mp) = \frac{(1\pm u)^2}{2U(1-\pm u/4)}\;.
\end{equation}

\section{Critical $Un$ for adiabatic invariance of $M$ from Bogoliubov stability analysis of the maximally excited array}

The calculation we perform to fix $u_c$ for the onset of $M$ decay is actually linear stability analysis of the maximally excited array state $\alpha_{1,r}=-\alpha_{2,r}=\sqrt{n/2}$. Since this maximally excited state has maximum $M$ value, its instability necessarily involves decrease of $M$. The reason that the stability of this very particular homogeneous state can be used to diagnose the invariance of $M$ for generic states is that the instability which we will find for the homogeneous state will turn out to be at the \emph{shortest} possible wavelengths. It is thus actually a local instability which can appear whenever even a few neighboring dimers are maximally excited. We have seen in our main text that the cases in which $M$ is adiabatically invariant, and in which it matters that $M$ is adiabatically invariant, are cases with bimodal dimer ensembles in which maximally excited dimers are not uncommon. The calculation of $u_c$ from the maximally excited state of the whole array thus shows when energy bimodality due to adiabatic invariance of $M$ becomes self-inconsistent, because above this $u_c$ threshold invariance of $M$ will lead to clusters of maximally excited dimers in which $M$ will spontaneously decay.

To perform the linear stability analysis we begin with the Hamiltonian equations of motion according to (1) from our main text, which are the two-component discrete Gross-Pitaevskii (GP) nonlinear Schr\"odinger equation
\begin{align}
i\dot{\alpha}_{1,jk} &= -\frac{J}{2} \sum_{\langle r,r^{\prime}\rangle} \alpha_{1,r^{\prime}} - \frac{\alpha_{2,r}}{2}  + U|\alpha_{1,r}|^2 \alpha_{1,r} \\
i\dot{\alpha}_{2,r} &= -\frac{J}{2} \sum_{\langle r,r^{\prime}\rangle} \alpha_{2,r^{\prime}} - \frac{\alpha_{1,r}}{2}  + U|\alpha_{2,r}|^2 \alpha_{2,r} \;,
\end{align}
where $r$ is short for the site indices $j,k$ in the 2D array.

We then linearize the above equations around the array's homogeneous maximally excited state $\alpha_{1,r}=-\alpha_{2,r}=\sqrt{n}$, which has $i\dot{\alpha}_{\sigma,r}=\nu\alpha_{\sigma,r}$ for
\begin{equation}
\nu = \frac{Un+1-J}{2}\;.
\end{equation}
Since the background about which we linearize has discrete translation symmetry in the $L\times L$ torus, we assume a discrete plane wave Ansatz
\begin{equation}
\alpha_{\sigma,jk}(t) = e^{-i\nu t}\left((-1)^\sigma\sqrt{\frac{n}{2}} + a_\sigma e^{\frac{2\pi i}{L}(p_1 j + p_2 k)}e^{-i\omega t} + b^*_\sigma e^{-\frac{2\pi i}{L}(p_1 j + p_2 k)}e^{-i\omega t}\right)
\end{equation}
for integers (discrete two-dimensional wave numbers) $\mathbf{p} = (p_1,p_2)$, linear excitation frequency $\omega(\mathbf{p})$, and infinitesimal two-component amplitudes $a_\sigma$, $b_\sigma$. Writing $u=Un$ as in the main text, this yields the time-independent Bogoliubov-de Gennes (BdG) equations
\begin{equation}
\omega \left(\begin{matrix} a_1 \\ a_2 \\ b_1 \\ b_2\end{matrix}\right) = \left(\begin{matrix} u + \mathcal{J}_\mathbf{p}-2 -\nu & \frac{1}{2} & \frac{u}{2} & 0 \\ \frac{1}{2} & u + \mathcal{J}_\mathbf{p}-2 -\nu & 0 \frac{u}{2}  \\ -\frac{u}{2} & 0 & -(u + \mathcal{J}_\mathbf{p}-2 -\nu) & -\frac{1}{2} \\ 0 & -\frac{u}{2} & -\frac{1}{2} & -(u + \mathcal{J}_\mathbf{p} -2-\nu)\end{matrix}\right)\left(\begin{matrix} a_1 \\ a_2 \\ b_1 \\ b_2\end{matrix}\right)
\end{equation}
where $\mathcal{J}_\mathbf{p} = 2J[\sin^2(\pi p_1/L)+\sin^2(\pi p_2/L)]$. 

The eigenfrequencies $\omega(\mathbf{p})$ of the four-by-four matrix have the two branches $\pm \sqrt{\mathcal{J}_\mathbf{p}(\mathcal{J}_\mathbf{p} + u)}$, which is always real, and
\begin{equation}
\omega =\pm \sqrt{(1 - \mathcal{J}_\mathbf{p})(1 - u - \mathcal{J}_\mathbf{p})}
\end{equation} 
which is imaginary, indicating dynamical instability, when $1-u < \mathcal{J}_\mathbf{p} < 1$. Since $\mathcal{J}_\mathbf{p}$ can always be made smaller by choosing smaller $\mathbf{p}$, the requirement for $\omega$ to be real for \emph{all} $\mathbf{p}$ is the requirement that the largest possible value $\mathcal{J}_\mathbf{p}$ is still smaller than $1-u$. The largest possible value of $\mathcal{J}_\mathbf{p}$ occurs for maximum $p_1 = p_2 = L$, giving $\mathcal{J}_\mathbf{p}\to 4J$. Thus the onset of instability, which occurs at the shortest possible wavelengths in the array, is $4J > 1- u$, so that the instability threshold is $u_c = 1-4J$ as stated in our main text. For the numerical evolutions shown in our main text we have used $J = 0.05$, giving $u_c = 0.8$.

\begin{figure}[ht]
\centering
\includegraphics[clip=true,width=0.5\columnwidth]{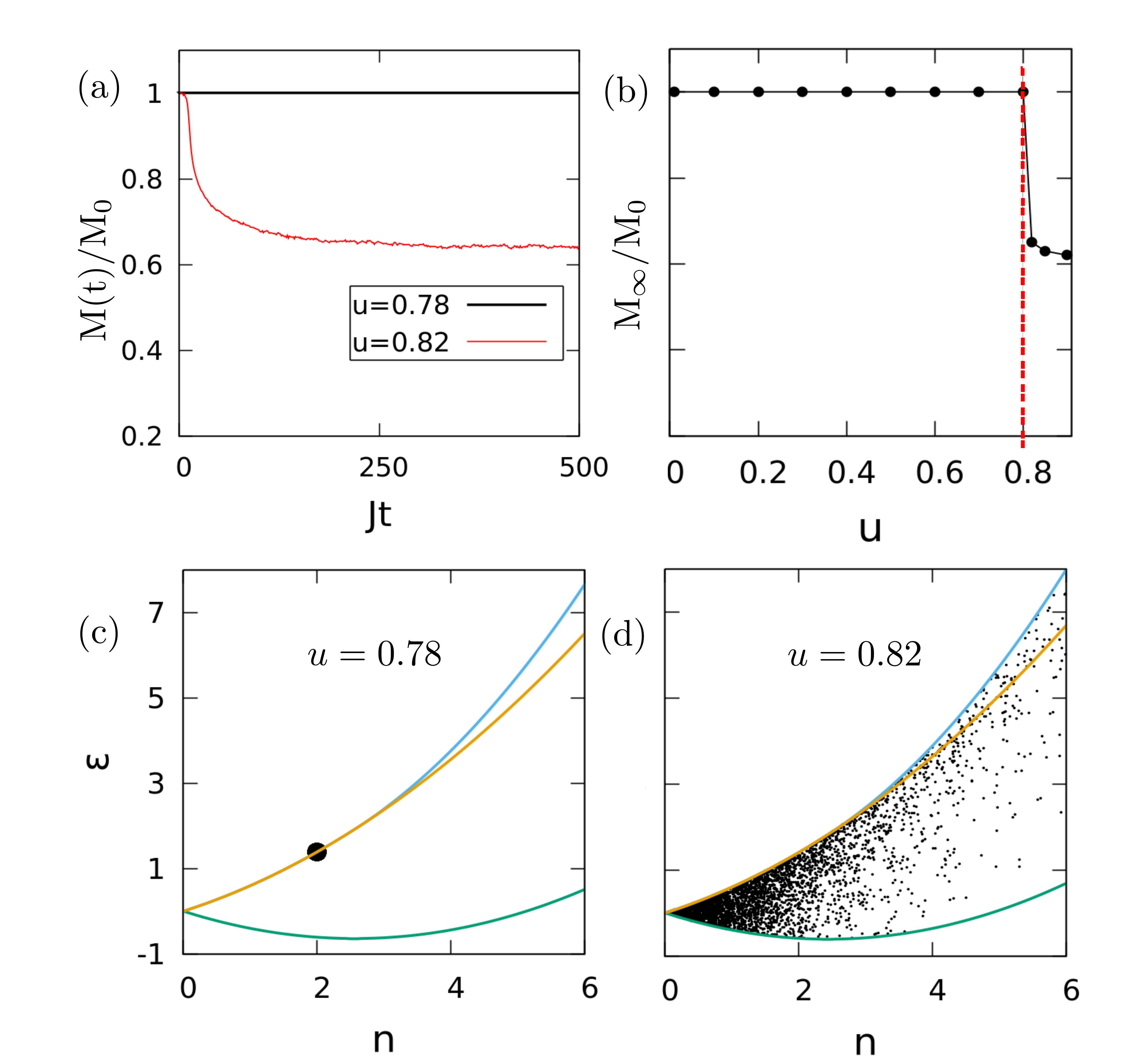}
\caption{(a) Dynamics of total dimer action variable $M$ for two representative values of $u$ below and above the critical $u_c$ for $J=0.05$. The corresponding $u_c$ is $0.8$ as marked by the vertical dotted line in (b), where the long time value $M_{\infty}$ is plotted against $u$. $M_0$ denotes the the total $M$ at $t=0$. (c-d) Final distribution of initially prepared maximally excited array of dimers in the $\varepsilon, n$ plane for $J=0.05$. Below $u_c$ the nearly uniform state remains stable, as seen in (c), while above $u_c$ it evolves into a dispersed ensemble of many different single-dimer configurations, scattered over the array.}
\label{joson_stability}
\end{figure}

We can confirm by numerical evolution of the array that the total dimer action $M=\sum_r m_r$ really does remain adiabatically invariant as long as the average $u<u_c$ but ceases to be invariant for average $u>u_c$. We do this by taking the initial state in which all the dimers of the array are at maximum energy ($\alpha_{2r}=-\alpha_{1r} = \sqrt{n/2}$), then adding a small random perturbation, and evolving over a long time. In Fig.\ \ref{joson_stability}a we have plotted the evolution of $M$ over time for two cases of average $u$, one just below $u_c$ and one just above it. The late-time value of $M$ is then plotted as a function of $u$ in Fig.\ \ref{joson_stability}b. The total $M$ is quite accurately conserved for $u$ below critical $u_c$, but for $u$ just above $u_c$ we see that $M$ drops substantially. (It does not necessarily decay to zero; total energy remains conserved and the single-dimer energy held in $m_r$ attains an equilibrium with the energy stored in long-wavelength sound waves in the Bose-Hubbard array.)  As explained above, the breakdown of adiabatic invariance of $M$ is due to short-wavelength instability in the array and so any small cluster of maximally excited dimers, such as tends to form if $M$ is invariant, will actually allow $M$ to decrease if $u> u_c$ in the cluster. 

\end{document}